\documentclass[aps,prstab,reprint,superscriptaddress]{revtex4-1}
\usepackage{graphicx}
\usepackage{booktabs}
\usepackage{amssymb}
\usepackage{graphics}
\usepackage{amsmath}
\usepackage{placeins}
\usepackage{hyperref}
\usepackage{verbatim}
\begin{document}

% Use the \preprint command to place your local institutional report
% number in the upper righthand corner of the title page in preprint mode.
% Multiple \preprint commands are allowed.
% Use the 'preprintnumbers' class option to override journal defaults
% to display numbers if necessary
%\preprint{}

%Title of paper
\title{Overestimation of thermal emittance in solenoid scans due to coupled transverse motion}

% repeat the \author .. \affiliation  etc. as needed
% \email, \thanks, \homepage, \altaffiliation all apply to the current
% author. Explanatory text should go in the []'s, actual e-mail
% address or url should go in the {}'s for \email and \homepage.
% Please use the appropriate macro foreach each type of information

% \affiliation command applies to all authors since the last
% \affiliation command. The \affiliation command should follow the
% other information
% \affiliation can be followed by \email, \homepage, \thanks as well.
\author{Lianmin Zheng}
%\homepage[]{Your web page}
%\thanks{}
%\altaffiliation{}
\affiliation{Department of Engineering Physics, Tsinghua University Beijing, Beijing 100084, People's Republic of China}
\affiliation{Key Laboratory of Particle and Radiation Imaging, Tsinghua University, Ministry of Education, Beijing 100084, People's Republic of China}
\affiliation{High Energy Physics Division, Argonne National Laboratory, Lemont, Illinois 60439, USA}
\author{Jiahang Shao}
\affiliation{High Energy Physics Division, Argonne National Laboratory, Lemont, Illinois 60439, USA}
\author{Yingchao Du}
\email[]{dych@mail.tsinghua.edu.cn}
\affiliation{Department of Engineering Physics, Tsinghua University Beijing, Beijing 100084, People's Republic of China}
\affiliation{Key Laboratory of Particle and Radiation Imaging, Tsinghua University, Ministry of Education, Beijing 100084, People's Republic of China}
\author{John G. Power}
\author{Eric E. Wisniewski}
\author{Wanming Liu}
\author{Charles E. Whiteford}
\author{Manoel Conde}
\author{Scott Doran}
\affiliation{High Energy Physics Division, Argonne National Laboratory, Lemont, Illinois 60439, USA}
\author{Chunguang Jing}
\affiliation{High Energy Physics Division, Argonne National Laboratory, Lemont, Illinois 60439, USA}
\affiliation{Euclid Techlabs LLC, Bolingbrook, Illinois 60440, USA}
\author{Chuanxiang Tang}
\author{Wei Gai}
\affiliation{Department of Engineering Physics, Tsinghua University Beijing, Beijing 100084, People's Republic of China}
\affiliation{Key Laboratory of Particle and Radiation Imaging, Tsinghua University, Ministry of Education, Beijing 100084, People's Republic of China}

%Collaboration name if desired (requires use of superscriptaddress
%option in \documentclass). \noaffiliation is required (may also be
%used with the \author command).
%\collaboration can be followed by \email, \homepage, \thanks as well.
%\collaboration{Jiahang Shao, John G. Power,}

%\noaffiliation

\date{\today}

\begin{abstract}
% insert abstract here
The solenoid scan is a widely used method for the in-situ measurement of the thermal emittance in a photocathode gun. The popularity of this method is due to its simplicity and convenience since all rf photocathode guns are equipped with an emittance compensation solenoid.  This paper shows that the solenoid scan measurement overestimates the thermal emittance in the ordinary measurement configuration due to a weak quadrupole field (present in either the rf gun or gun solenoid) followed by a rotation in the solenoid.  This coupled transverse dynamics aberration introduces a correlation between the beam's horizontal and vertical motion leading to an increase in the measured 2D transverse emittance, thus the overestimation of the thermal emittance.  This effect was systematically studied using both analytic expressions and numerical simulations. These studies were experimentally verified using an L-band 1.6-cell rf photocathode gun with a cesium telluride cathode, which shows a thermal emittance overestimation of 35\% with a rms laser spot size of 2.7~mm.  The paper concludes by showing that the accuracy of the solenoid scan can be improved by using a quadrupole magnet corrector, consisting of a pair of normal and skew quadrupole magnets.

\end{abstract}

% insert suggested PACS numbers in braces on next line
% insert suggested keywords - APS authors don't need to do this
%\keywords{}

%\maketitle must follow title, authors, abstract, \pacs, and \keywords
\maketitle

% body of paper here - Use proper section commands
% References should be done using the \cite, \ref, and \label commands
\section{INTRODUCTION}\label{intro}

\par The solenoid scan is one of the most commonly used methods for the in-situ measurement of the thermal emittance of a photocathode in an electron gun. \cite{bazarov2008thermal,hauri2010intrinsic,qian2012experimental,lee2015review,maxson2017direct,graves2001measurement,gulliford2013demonstration,miltchev2005measurements,bazarov2011thermal}. The measurement has a simple experimental configuration: an rf or dc photocathode gun followed by a transport line consisting of a solenoid and a drift. The photoelectron beam exits the gun at relatively high energy after which it immediately enters the transport line where it is focused by a solenoid onto a screen located at the end of a drift. The thermal emittance of the photocathode is obtained by measuring the beam's transverse size on the screen as a function of the solenoid focusing strength and then fitting these sizes according to the linear transfer matrix of the transport line. 

\par Driven by the desire for high brightness electron sources, the thermal emittance of both metal and semiconductor cathodes has been intensively investigated with measurements in the past few decades \cite{hauri2010intrinsic,qian2012experimental,graves2001measurement,gulliford2013demonstration,miltchev2005measurements,prat2015measurements}. Some of these measurements have deviated significantly from the theoretical predictions. For copper cathodes illuminated by a 266~nm laser, the theoretical thermal emittance is 0.5~mm\,mrad/mm \cite{dowell2010cathode} while the measured values vary from 0.57~mm\,mrad/mm \cite{prat2015measurements} to 1.17~mm\,mrad/mm \cite{qian2012experimental}. For cesium telluride cathodes and a 262~nm laser, the thermal emittance is predicted to be 0.9~mm\,mrad/mm \cite{dowell2010cathode} while the measured values vary from 0.54~mm\,mrad/mm \cite{prat2015measurements} to 1.2~mm\,mrad/mm \cite{miltchev2005measurements}. These discrepancies are only partially explained by actual increases of the thermal emittance (due to surface roughness and/or impurities) \cite{qian2012experimental,zhang2015analytical} so the remainder of the discrepancies must be due to the measurement method itself.  

\par Accurate measurement of the thermal emittance via the solenoid scan method depends on three factors: (1) accurate beam size measurement at the screen (2) accurate knowledge of the transfer matrix and (3) reduction of the sources of emittance growth so that only the thermal emittance remains.  The first factor can be improved by employing a high sensitivity CCD camera \cite{qian2012experimental} and a thin YAG:Ce screen \cite{maxson2017direct}.  The second factor requires accurate knowledge of the fields of the beamline elements and the distances between the elements.  Finally, in the third category, there are a number of well-known factors that increase the emittance of the beam thus leading to an overestimation of the thermal emittance.  Inside the gun, these known factors include the nonlinear effects from space charge (SC) \cite{lee2015review,miltchev2005measurements} and rf effects \cite{chae2011emittance} that increase the projected emittance.  SC effects are mitigated by using low charge beams while rf effects are reduced by using short beams.  After the gun, the solenoid’s spherical and chromatic aberrations \cite{dowell2016sources, mcdonald1989methods} will also induce growth of the rms emittance.  Both of these aberrations scale with the square of the transverse beam size.  In theory, these effects can be mitigated by keeping the beam size small inside the bore of the solenoid but in practice the beam often becomes large inside the solenoid during the scan.  The chromatic aberration scales with the beam's energy spread and can therefore be mitigated by operating with a short bunch at low charge - consistent with mitigation of rf and SC effects in the gun.  In summary, the traditional solenoid scan method uses a bunch with low charge and short length to reduce the SC, rf, and chromatic sources of emittance growth.  
 
\par  This paper presents a previously overlooked source of emittance growth due to coupled transverse dynamics aberration that leads to an overestimation of the thermal emittance measured via the solenoid scan.  The work presented in this paper was inspired by the recent publication by Dowell et al. \cite{,dowell2018exact}.  Whereas the reference focused on the sources of emittance growth due to the coupled transverse dynamics aberration and its elimination, here we are focused on the impact of this aberration on the thermal emittance measurement via the solenoid scan which arises inescapably because the scanning solenoid itself is the source of the aberration.  This situation was not addressed in Ref.~\cite{dowell2018exact} so this paper presents a systematic study of the thermal emittance overestimation from the coupled transverse dynamics aberrations using the solenoid scan technique.  This aberration arises when the beam motion in the $x-x'$ plane becomes correlated with $y-y'$ plane causing an emittance growth in the 2D phase space. Two aberration sources (thoroughly explained in the reference \cite{,dowell2018exact}) exist in the measurement beamline: (1) an rf quadrupole field, in the rf gun, followed by a rotation in the solenoid, and (2) a constant quadrupole field, inside or before the solenoid, followed by a rotation in the solenoid.   In the remainder of this paper, these are referred to as the gun quadrupole and the solenoid quadrupole.

\par This paper is organized as follows. Section \ref{section2} describes the gun and solenoid quadrupole focusing in the solenoid scan beamline. Section \ref{section3} discusses the emittance growth due to the coupled transverse dynamics aberration in the solenoid scan beamline. In section \ref{s3}, the thermal emittance overestimation due to the aberrations in the solenoid scan technique is studied both analytically and numerically. In section \ref{ex}, the overestimation is experimentally verified using an L-band 1.6-cell photocathode rf gun with a cesium telluride cathode. Finally, we propose a flexible and compact quadrupole corrector in section \ref{corrector} to minimize the coupled transverse dynamics aberrations so as to improve the accuracy of thermal emittance measurement using solenoid scan.

\section{Quadrupole focusing in the solenoid scan beamline}\label{section2}

\par
\begin{figure}[hbtp]
	\centering
	\includegraphics[width=0.43\textwidth]{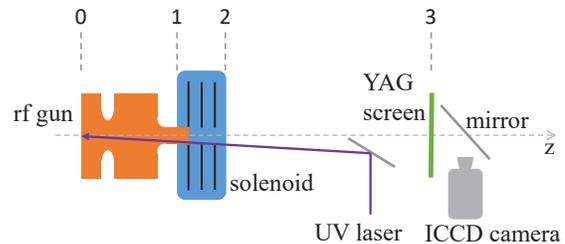}
	\caption{\label{FIG.beamline_show} Thermal emittance measurement setup at AWA.}
\end{figure}

 In this section we derive transfer matrices for the quadrupole focusing that arises in the solenoid scan beamline.  The layout of the beamline is shown in FIG.~\ref{FIG.beamline_show}. Here the cathode of the rf gun, the solenoid entrance, the solenoid exit and the YAG screen are marked as Position 0 to 3 respectively. In this section only the beamline from the cathode (Position 0) to the solenoid exit (Position 2) are used. Undesired quadrupole fields often exist in rf photocathode guns and, to the best of our knowledge, always exist in the solenoid field of all fabricated solenoid magnets used with rf guns.  The quadrupole fields can be aligned either normally or rotated about the z-axis (beam transport direction).   The focusing due to quadrupole components is presented for three cases: the rf gun alone, the solenoid alone, and simultaneously in both the gun and solenoid. 

\subsection{quadrupole focusing from the rf gun}

\par
\begin{figure}[hbtp]
	\centering
	\includegraphics[width=0.4\textwidth]{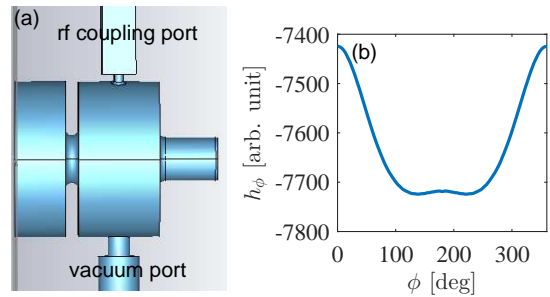}
	\caption{\label{FIG.gun_image} The drive gun at the AWA. (a) 3D model in CST Microwave Studio; (b) Azimuthal magnetic field $h_{\phi}$ along the angular direction at the center of the full cell with a 15~mm radius. A Fourier analysis shows that the quadrupole strength with respect to the monopole one is 7.2$\times10^{-3}$, which is much larger than the LCLS gun (1.2$\times10^{-4}$) \cite{houjunphdthesis} and the newly designed Tsinghua gun (5.6$\times10^{-5}$) \cite{houjunphdthesis}.}
\end{figure}

 Quadrupole fields often exist in rf photocathode guns due to the asymmetric geometry of the cells due to openings in the side walls: rf coupling ports, pumping ports, laser ports, etc. Recent rf guns have eliminated the quadrupole field by using racetrack \cite{Xiao2005Dual} or four-port \cite{hong2011high,zheng2016development} geometries in the cells. However, many older rf guns, without these symmetrizing features, are still in operation since the redesign, fabrication, and commissioning of a new gun is time-consuming and expensive. The gun used in our study (the drive gun \cite{conde2001argonne} at the Argonne Wakefield Accelerator (AWA) facility) is of the older style and has only one rf coupling port and a small vacuum port on the opposite side of the cell, as illustrated in FIG.~\ref{FIG.gun_image}(a). This gun has a strong quadrupole field, as illustrated in the CST Microwave Studio simulation \cite{CST} shown in FIG.~\ref{FIG.gun_image}(b).

\par The AWA rf drive gun has a normal quadrupole component due to the location of its rf coupling port and vacuum pumping port in the vertical direction. The normalized transverse momentum due to the gun quadrupole can be expressed as\cite{chae2011emittance}
\begin{equation}\label{kick}
\frac{p_ \bot }{p_z} = -2a\alpha L\sin {\varphi _0}(x\hat x - y\hat y)
\end{equation}
where $p_ \bot$ is the transverse momentum and $p_z$ is the longitudinal momentum. $a$ is the parameter characterizing the relative strength of the quadrupole field to the monopole one, $\alpha$ is the normalized rf field strength, $L$ is the full cell length, and $\varphi _0$ is the phase when the electron arrives at the full cell entrance.
\par The beam trajectory in the trace space $x$ and $x'$ after the gun quadrupole is given by 
\begin{equation}\label{trajectory}
\left[ {\begin{array}{*{20}{c}}
x\\
{x'}
\end{array}} \right] = \left[ {\begin{array}{*{20}{c}}
1&0\\
{ - \frac{1}{{{f_g}}}}&1
\end{array}} \right]\left[ {\begin{array}{*{20}{c}}
{{x_0}}\\
{{{x'}_0}}
\end{array}} \right]
\end{equation}
where ${f_g} $ is the equivalent focal length of the gun quadrupole.

\par Based on Eqn.~(\ref{kick}) and (\ref{trajectory}) and assuming $x'_0=0$, the slope of the trajectory after the gun quadrupole should be $x' = \frac{{{p_x}}}{{{p_{z}}}} =  - 2a\alpha L\sin {\varphi _0}{x_0} =  - \frac{{{x_0}}}{{{f_g}}}$. Therefore, the focal length due to the gun quadrupole is
\begin{equation}\label{fgun}
{f_g} = \frac{1}{{2a\alpha L\sin {\varphi _0}}}
\end{equation}
and the transfer matrix due to the normal quadrupole focusing in the rf gun can be expressed as
\begin{equation}
{R_{gunquad}} = \left[ {\begin{array}{*{20}{c}}
1&0&0&0\\
{-\frac{1}{{{f_g}}}}&1&0&0\\
0&0&1&0\\
0&0&{\frac{1}{{{f_g}}}}&1
\end{array}} \right]
\end{equation}

\subsection{quadrupole focusing from the solenoid}

\par The AWA solenoid has quadrupole fields due to the asymmetry of the solenoid's yoke and/or coil windings.   While its quadrupole field has not been measured via a rotating wire, it should be similar in character to the LCLS solenoid shown in FIG.~2 of Ref.~\cite{dowell2018exact} where the solenoid field was measured to have a rotated quadrupole component at the entrance and exit of the solenoid.  (This assumption is validated in Section~\ref{ex} where the angle of the rotated quadrupole is measured using a beam-based method.)  We only need to consider the quadrupole located at the solenoid entrance since the focusing followed by the rotation is the source of the coupling between the transverse planes as explained above.  Let the rotated quadrupole field in the solenoid have strength ${f_s}$ and rotation angle $\eta$, then its transfer matrix can be written as
\begin{equation}
{R_{solquad}} = \left[ {\begin{array}{*{20}{c}}
1&0&0&0\\
{ - \frac{{\cos 2\eta }}{{{f_s}}}}&1&{ - \frac{{\sin 2\eta }}{{{f_s}}}}&0\\
0&0&1&0\\
{ - \frac{{\sin 2\eta }}{{{f_s}}}}&0&{\frac{{\cos 2\eta }}{{{f_s}}}}&1
\end{array}} \right]
\end{equation}

\subsection{combined quadrupole focusing from the rf gun and solenoid}

\par  When both quadrupole fields are present, the above two transfer matrices of the gun and the solenoid quadrupoles can be easily combined as
\begin{equation}\label{zuhe}
\begin{aligned}
{R_{(g + s)quad}} =& \left[ {\begin{array}{*{20}{c}}
1&0&0&0\\
{ - \frac{{\cos 2\eta }}{{{f_s}}}}&1&{ - \frac{{\sin 2\eta }}{{{f_s}}}}&0\\
0&0&1&0\\
{ - \frac{{\sin 2\eta }}{{{f_s}}}}&0&{\frac{{\cos 2\eta }}{{{f_s}}}}&1
\end{array}} \right]\left[ {\begin{array}{*{20}{c}}
1&0&0&0\\
{ - \frac{1}{{{f_g}}}}&1&0&0\\
0&0&1&0\\
0&0&{\frac{1}{{{f_g}}}}&1
\end{array}} \right]\\
 =& \left[ {\begin{array}{*{20}{c}}
1&0&0&0\\
{ - \frac{1}{{{f_g}}} - \frac{{\cos 2\eta }}{{{f_s}}}}&1&{ - \frac{{\sin 2\eta }}{{{f_s}}}}&0\\
0&0&1&0\\
{ - \frac{{\sin 2\eta }}{{{f_s}}}}&0&{\frac{{\cos 2\eta }}{{{f_s}}} + \frac{1}{{{f_g}}}}&1
\end{array}} \right]\\
 =& \left[ {\begin{array}{*{20}{c}}
1&0&0&0\\
{ - \frac{{\cos 2\theta }}{{{f_c}}}}&1&{ - \frac{{\sin 2\theta }}{{{f_c}}}}&0\\
0&0&1&0\\
{ - \frac{{\sin 2\theta }}{{{f_c}}}}&0&{\frac{{\cos 2\theta }}{{{f_c}}}}&1
\end{array}} \right]
\end{aligned}
\end{equation}
where the combined focusing strength $f_c$ and rotation angle $\theta$ of the combined transfer matrix (Eqn.~\ref{zuhe}) is given by

\begin{equation}\label{combined}
	\left\{
	\begin{aligned}
		{f_c} &= \frac{1}{{\sqrt {\frac{1}{{{f_g}^2}} + \frac{{2\cos 2\eta }}{{{f_g}{f_s}}} + \frac{1}{{{f_s}^2}}} }}\\
		\theta &= \frac{1}{2}\arcsin \left( {\frac{{{f_c}}}{{{f_s}}}\sin 2\eta } \right)\\
	\end{aligned}
	\right.
\end{equation}

\section{Emittance growth due to the coupled aberrations in the solenoid scan beamline}\label{section3}
\par In this section, we present an analytical estimate of the emittance growth due to the coupled transverse dynamics aberration present in the solenoid scan and verify the estimate with numerical simulations.  The transverse coupling aberration in the solenoid scan beamline is generated when the quadrupole focusing first focuses the beam to an elliptical shape, the beam is rotated in the solenoid which results in the transverse coupling. We analyze the emittance growth for the same three cases outlined in the previous section; once again only the beamline from the cathode (Position 0) to the solenoid exit (Position 2) is used here.
\par The emittance of the beam, after passing through the gun and solenoid, is given by
\begin{equation}\label{total_emittacne}
{\varepsilon} = \sqrt {{\varepsilon _{therm}}^2 + {\varepsilon _{coupled}}^2 + {\varepsilon _{other}}^2} 
\end{equation}
where $\varepsilon _{therm}$ is the thermal emittance, $\varepsilon _{coupled}$ is the emittance growth due to the transverse coupled dynamics, and $\varepsilon _{other}$ is the emittance growth due to space charge, rf, spherical and chromatic aberrations, etc.  Therefore, to estimate $\varepsilon$ we need separate estimates of its three components.  In the ideal solenoid scan case the final emittance is equal to the thermal emittance so the second and third terms of Eqn.~\ref{total_emittacne} should be zero.  This is true for the last term, $\varepsilon _{other}$, since the solenoid scan parameters are chosen to minimize it as described in section~\ref{intro}, so $\varepsilon _{other}\approx0$ for our analytical estimate.  However, we show below that the middle term is, in general, not zero and this term causes a growth of the final emittance.
\par The thermal emittance $\varepsilon _{therm}$, is estimated with the three-step model.  It can be expressed as $\epsilon=\sigma_l\sqrt{\frac{2E_K}{3m_ec^2}}$, where $\sigma_l$ is the rms laser spot size and $m_ec^2$ is the electron rest energy~\cite{flottmann1997note}. The excess energy of the cesium telluride cathode can be expressed as $2E_K=\phi_l-E_g-E_a+\phi_{Sch}$, where $\phi_l$ is the photon energy, $E_g$ is the gap energy, $E_a$ is the electron affinity, $\phi_{Sch}=\sqrt{\frac{e^3}{4\pi\epsilon_0}\beta E_c}$ is the barrier reduction by the applied electric field due to the Schottky effect \cite{chen2012surface}. The typical cathode barrier $E_g+E_a$ of the cesium telluride is reported to be 3.5~eV \cite{dowell2010cathode,miltchev2005measurements}. By assuming the field enhancement factor $\beta=1$ and using 248~nm UV laser ($\phi_l=5$~eV), the theoretical thermal emittance should be 1.05~mm\,mrad/mm.   In the numerical simulation and during the experiment, the initial electron beam spot size has a uniform transverse distribution with 12~mm diameter (rms spot size 3~mm).  Therefore, the estimated rms thermal emittance is 3~mm$\times$(1.05~mm\,mrad/mm), or 3.15~mm mrad.
\par The coupled emittance estimate after passing through the solenoid, according to Ref.~\cite{dowell2018exact}, is given by
\begin{equation}\label{emittacne_coupled}
{\varepsilon _{coupled}} = \beta \gamma \frac{{{\sigma _{x,sol}}{\sigma _{y,sol}}}}{{{f_c}}}\left| {\sin 2(KL + \theta )} \right|
\end{equation}
where $K = \frac{eB_0}{2\beta\gamma mc}$, $L$, $KL$, and $B_0$,  denote the strength, the effective length, the Larmor angle, and the peak magnetic field of the solenoid, respectively.    Therefore, the analytical estimate of $\varepsilon$ has a minimum value of  $3.15~mm mrad$ due to $\varepsilon _{therm}$ added in quadrature to the sinusoidal oscillation term of $\varepsilon _{coupled}$ given in Eqn.~\ref{emittacne_coupled}. 

\par To verify the analytic estimate of the emittance growth, an ASTRA \cite{floettmann2011astra} beam dynamics simulation was performed.  In the simulation results below, the cathode gradient is 32~MV/m, corresponding to a maximum acceleration phase of 37$^{\circ}$ and an ideal solenoid is used (i.e. one without quadrupole focusing) with peak field is fixed at 0.1974~T corresponding to a Larmor angle $KL$ of -30$^{\circ}$.
\par To minimize $\varepsilon _{other}$ in the ASTRA simulation, a short pulse, low charge bunch is used.  The rf emittance growth due to the phase-dependent rf kick (including dipole, quadrupole and higher order fields) \cite{chae2011emittance} was minimized to 1.4\% by the use of a 1.5~ps FWHM pulse length and verified by ASTRA simulations.  The emittance growth due to space charge is zero since the charge is set to zero during the ASTRA simulations.  This was done to speed up the simulations but we also confirmed that a sub-picocoulomb charge has less than 1\% contribution to the final emittance.  Note that this short pulse, low charge combination also minimizes the emittance growth due to the chromatic aberration of the solenoid to 1.6\%. 

\subsection{emittance growth due to the gun quadrupole}
\par
\begin{figure}[hbtp]
\centering
\includegraphics[width=0.4\textwidth]{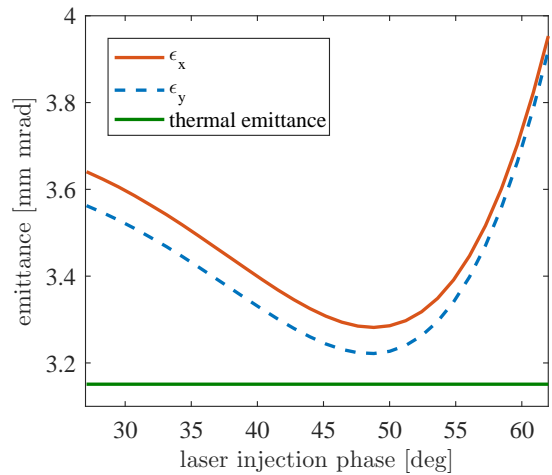}
\caption{\label{FIG.emt_phase} Emittance at Position 2 based on the realistic rf gun (includes quadrupole) and ideal solenoid (no quadrupole component) at different laser injection phases.  The gun launch phase corresponding to $\varphi_0=0^\circ$ is 49${^\circ}$.}
\end{figure}

 For the first case, only the gun quadrupole is taken into consideration while ignoring the one in the solenoid.  In the ASTRA simulation, a 3D rf field map was used for the gun (exported from CST Microwave Studio) and an ideal 1D field map (exported from POISSON) for the solenoid.  The total simulated emittance after the ideal solenoid as a function of the laser injection phase is shown in Figure ~\ref{FIG.emt_phase}.  This simulation result can be compared to the analytic one by setting $f_c=f_g$, $f_s=\infty$, $\varepsilon _{therm}=3.15~$mm mrad.  The emittance growth due to the coupled aberration can be found by substituting Eqn.~\ref{fgun} into Eqn.~\ref{emittacne_coupled} which will be zero when $\varphi_0=0^{\circ}$.  This corresponds to a laser injection phase of 49$^{\circ}$ which was found by simulating the electron travel time from the cathode to the full cell.  The minimum total emittance is close to the thermal emittance value (3.15~mm mrad) when the laser injection phase is 49${^\circ}$, which demonstrates good agreement between the simulation and analytic results. The difference between the minimal total emittance in the ASTRA simulation and the thermal value is mainly caused by the aforementioned rf effects and chromatic effects.

\subsection{emittance growth due to the solenoid quadrupole}
\begin{figure}[hbtp]
\centering
\includegraphics[width=0.4\textwidth]{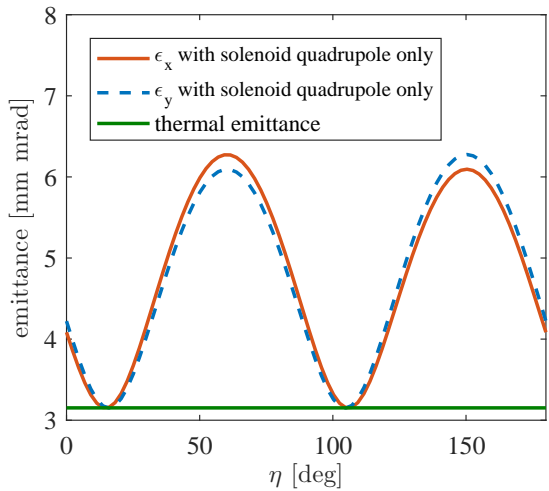}
\caption{\label{FIG.emt_eta_solenoid_only} Emittance at Position 2 based on the ideal rf gun (no quadrupole component) and realistic solenoid (includes quadrupole) as a function of the rotation angle of the solenoid quadrupole.}
\end{figure}

For the second case, only the solenoid quadrupole is taken into consideration while ignoring the one in the gun. In the ASTRA simulation, a 1D rf field map in the gun is used instead of the 3D field map and, once again, an ideal 1D field map (exported from POISSON) for the solenoid. To model the solenoid quadrupole, a quadrupole element was added to ASTRA at the same location as the solenoid.  Its longitudinal field profile is the same as the ideal solenoid and its strength is set to 77~Gauss/m based on the experimental study introduced in Sec.~\ref{ex}. The simulated emittance after the solenoid/quad location (Position 2) as a function of the rotation angle of the solenoid quadrupole $\eta$  is shown in FIG.~\ref{FIG.emt_eta_solenoid_only}.  Comparing this simulation result to the analytic expression in Eqn.~(\ref{emittacne_coupled}) (by setting $f_c=f_s$, $f_g=\infty$, $\varepsilon _{therm}=3.15~\rm mm mrad$) we see the emittance oscillates sinusoidally with $\theta=\eta$ which again demonstrates good agreement between the simulation and analytic results.

\subsection{emittance growth due to both the gun and in the solenoid quadrupole}
\begin{figure}[hbtp]
\centering
\includegraphics[width=0.4\textwidth]{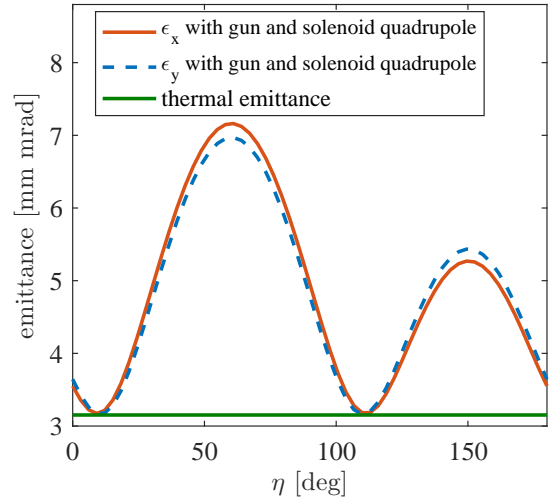}
\caption{\label{FIG.emt_eta_gun_solenoid} Emittance at Position 2 based on the realistic rf gun (includes quadrupole) and solenoid (includes quadrupole) as a function of the rotation angle of the solenoid quadrupole.}
\end{figure}

For the third and final case, both the gun quadrupole and the solenoid quadrupole are taken into consideration. In the ASTRA simulation, a 3D rf field map was used for the gun (exported from CST Microwave Studio), an ideal 1D field map (exported from POISSON) for the solenoid and a quadrupole element was added to the beamline as described above. The laser injector phase is fixed at 43 deg which produces negative $\varphi_0=-6^\circ$ and $f_g=-137$ m. The total simulated emittance after the solenoid/quad location (Position 2) as a function of the rotation angle of the solenoid quadrupole $\eta$  is shown in FIG.~\ref{FIG.emt_eta_gun_solenoid}. Note that the emittance oscillation curves are different in comparison to FIG.~\ref{FIG.emt_eta_solenoid_only} due to the combined focal length $f_c$ and the angle $\theta$. The gun quadrupole can partially cancel or add to the solenoid quadrupole when $\eta$ is around $150^\circ$ or $60^\circ$, making the emittance growth smaller or larger. 

\section{Thermal emittance overestimation in solenoid scan}\label{s3}
\par In this section we show that the emittance measured by the solenoid scan, i.e., the fitted emittance based on the fitting of the rms beamsize and the solenoid strength, $\varepsilon _{fit}$, is an overestimation of the thermal emittance.  Further, $\varepsilon _{fit}$, is approximately equal to the quadrature sum of the thermal emittance,  $\varepsilon _{therm}$, and the emittance growth due to the transverse coupled aberration, $\varepsilon _{coupled}$.  

\subsection{solenoid scan formalism}\label{formalism}
\par First we present the formalism for the normal solenoid scan, i.e. without any aberrations. Similar to the previous section, $\varepsilon _{other}\approx0$ since the solenoid scan parameters are chosen to minimize it as described in section~\ref{intro}.  The solenoid scan beamline begins at the solenoid entrance and ends at the YAG screen; i.e. from positions $1\rightarrow3$ (FIG.~\ref{FIG.beamline_show}).  Its transfer matrix is given by the linear transfer matrix for the solenoid $R_{sol}$ and the drift $R_d$ successively (quadrupole components are not considered in the normal solenoid scan). The solenoid's matrix can be expressed as

\begin{equation}\label{sol_matrix}
{R_{sol}} = \left[ {\begin{array}{*{20}{c}}
	{{C^2}}&{\frac{{SC}}{K}}&{SC}&{\frac{{{S^2}}}{K}}\\
	{ - KSC}&{{C^2}}&{ - K{S^2}}&{SC}\\
	{ - SC}&{ - \frac{{{S^2}}}{K}}&{{C^2}}&{\frac{{SC}}{K}}\\
	{K{S^2}}&{ - SC}&{ - KSC}&{{C^2}}
	\end{array}} \right]=R_{rot}R_{foc}
\end{equation}
where $S\equiv{\rm sin}(KL)$, $C\equiv{\rm cos}(KL)$. $R_{rot}$ and $R_{foc}$ are the rotation matrix and the focusing matrix respectively.
\begin{equation}
{R_{rot}} = \left[ {\begin{array}{*{20}{c}}
C&0&S&0\\
0&C&0&S\\
{ - S}&0&C&0\\
0&{ - S}&0&C
\end{array}} \right]
\end{equation}
\begin{equation}
{R_{foc}} = \left[ {\begin{array}{*{20}{c}}
C&{\frac{S}{K}}&0&0\\
{ - KS}&C&0&0\\
0&0&C&{\frac{S}{K}}\\
0&0&{ - KS}&C
\end{array}} \right]
\end{equation}

The drift's matrix can be expressed as
\begin{equation}\label{R_d}
{R_d} = \left[ {\begin{array}{*{20}{c}}
	1&{{L_d}}&0&0\\
	0&1&0&0\\
	0&0&1&{{L_d}}\\
	0&0&0&1
	\end{array}} \right]
\end{equation}
where $L_{d}$ is the length of the drift. 

The thermal emittance $\epsilon_x$ and $\epsilon_y$ are complicated to deduce because the beam trajectories in x and y directions are coupled due to the rotation matrix of the solenoid. For simplicity the rotation term $R_{rot}$ is usually ignored in the beam moments calculation \cite{qian2012experimental,lee2015review,Scifo2018}, and the transfer matrix of the solenoid scan beamline without aberrations is expressed as $R\equiv R_d R_{foc}$. Therefore, the beam size squared at the end of the drift is
\begin{equation}\label{eq10}
\begin{aligned}
{\sigma _{3}}^2  =& {R(1,1)}^2\langle {x_1}^2\rangle  + 2{R(1,1)}{R(1,2)}\langle {x_1}{x'_1}\rangle\\
&+{R(1,2)}^2\langle {x'_1}^2\rangle 
\end{aligned}
\end{equation}
where $\langle {x_1}^2\rangle$, $\langle {x_1}{{x'}_1}\rangle$, and $\langle {{x'}_1}^2\rangle$ are the beam moments at the solenoid entrance. Note that the solenoid scan method requires prior knowledge of $R$ to find the beam moments and thus the emittance. Substituting in the values from the transfer matrices we obtain an analytical expression for the expected x-beam spot size squared at 3,
\begin{equation}\label{eq33a}
\begin{aligned}
{\sigma _{3}}^2 =& {\left( {C - {L_d}KS} \right)^2}\langle {x_{1}}^2\rangle \\
&  + 2\left( {C - {L_d}KS} \right) \left( {S/K + C{L_d}} \right)\langle {x_{1}}{{x'}_{1}}\rangle \\
&  + {\left( {S/K + C{L_d}} \right)^2}\langle {{x'}_{1}}^2\rangle \\ 
\end{aligned}
\end{equation}
\subsection{Measured spot sizes on the screen}
\par  In this section, we simulate the measured spot sizes on the screen by propagating the initial beam moments (Position 1 in FIG.~\ref{FIG.beamline_show}) through the solenoid scan beamline transfer matrix $M$ (which now includes aberrations) to the YAG screen (Position 3) for different solenoid strength settings $K$. This beamline is the same as the one in Sec.~\ref{formalism} except that a thin quadrupole lens is now placed at the solenoid entrance (Position 1).  For simplicity, let the thin quadrupole lens have a normal rotation angle with focal length of $f$, so its transfer matrix can be expressed as 
\begin{equation}\label{R_f1}
{R_f} = \left[ {\begin{array}{*{20}{c}}
1&0&0&0\\
{ - \frac{1}{f}}&1&0&0\\
0&0&1&0\\
0&0&{\frac{1}{f}}&1
\end{array}} \right]
\end{equation}
and the transfer matrix of the solenoid scan beamline with aberrations is $M \equiv R_d R_{sol} R_f$.

\par The initial beam is characterized by the beam sigma matrix at the solenoid entrance (Position 1 in FIG.~\ref{FIG.beamline_show}). In order to keep the analysis simple, we assume the initial beam has zero-emittance, uniform transverse distribution, with the same beam size squared in both x and y directions ($\langle {x_1}^2\rangle$), and perfectly parallel rays.  The initial beam matrix and initial emittance at the solenoid entrance (Position 1 in FIG.~\ref{FIG.beamline_show}) can be expressed as
\begin{equation}\label{eq150}
\begin{array}{l}
{\Sigma _1} = \left( {\begin{array}{*{20}{c}}
	{\langle {x_1}^2\rangle}&0&0&0\\
	0&0&0&0\\
	0&0&{\langle {x_1}^2\rangle}&0\\
	0&0&0&0
	\end{array}} \right) \\
 \varepsilon_{1}=0
\end{array}
\end{equation}
Eqn.~\ref{eq150} completely specifies the initial beam conditions. 
\par The beam sigma matrix at the screen (Position 3 in FIG.~\ref{FIG.beamline_show}) can be expressed as
\begin{equation}\label{eq31}
{\Sigma _{scr}} = M{\Sigma _1}{M^T}
\end{equation}
so that the measured x-beam sizes squared at the screen $\sigma_{scr}^2$ as a function of the solenoid strength $K$ can be expressed as
\begin{equation}\label{eq100}
\begin{array}{l}
{\sigma _{scr}}^2 = {\Sigma _{scr}}(1,1) = \\
\frac{\langle {x_1}^2\rangle}{{{f^2}{K^2}}}\left[ \begin{array}{l}
{\left( {K(f + {L_d})SC + (1 - f{K^2}{L_d}){S^2}} \right)^2} + \\
{\left( {K(f - {L_d}){C^2} - (1 + f{K^2}{L_d})SC} \right)^2}
\end{array} \right]
\end{array}
\end{equation}

\subsection{Fitting}
\par To retrieve the emittance measured by the solenoid scan, the x-spot sizes squared at the screen (Eqn.~\ref{eq100}) are compared to the analytical expectation of the beam sizes squared (Eqn.~\ref{eq33a}) in order to obtain the fitted beam moments at Position 1: $\langle {x_{fit}}^2\rangle$, $\langle {x_{fit}}{{x'}_{fit}}\rangle$ and $\langle {{x'}_{fit}}^2\rangle$.  Note that if the solenoid scan beamline had no aberrations then the fitting routine would retrieve the initial beam moments (Eqn.~\ref{eq150}); i.e. $\langle {x_{fit}}^2\rangle=\langle {x_{1}}^2\rangle$,  $\langle {x_{fit}}{{x'}_{fit}}\rangle=\langle {{x'}_{fit}}^2\rangle=0$, and the measured (fitted) emittance $\varepsilon_{fit}=0$.  However, as we will show this is not the case due to the transverse coupled aberration.  The fitted beam sizes squared ${\sigma _{fit}}^2$ as a function of the fitted beam moments can be expressed as 
\begin{equation}\label{eq33}
\begin{aligned}
{\sigma _{fit}}^2 =& {\left( {C - {L_d}KS} \right)^2}\langle {x_{fit}}^2\rangle \\
&  + 2\left( {C - {L_d}KS} \right) \left( {S/K + C{L_d}} \right)\langle {x_{fit}}{{x'}_{fit}}\rangle \\
&  + {\left( {S/K + C{L_d}} \right)^2}\langle {{x'}_{fit}}^2\rangle \\ 
\end{aligned}
\end{equation}
and the goal of the fitting routine is to minimize $\left| {{\sigma _{fit}}^2 - {\sigma _{scr}}^2} \right|$ to retrieve the fitted beam moments $\langle {x_{fit}}^2\rangle$, $\langle {x_{fit}}{{x'}_{fit}}\rangle$ and $ \langle {{x'}_{fit}}^2\rangle$ and thus the fitted emittance.

\par The next step is to scan the solenoid strength $K$ in order to generate a series of spot sizes on the screen $\sigma_{scr}^2$ (Eqn.~\ref{eq100}) and then fit them to $\sigma_{fit}^2$ (Eqn.~\ref{eq33}).  A Taylor expansion method is used to scan $K$ about its value at the beam waist ($k_0$). During the solenoid scan, the maximum beam size at the screen is typically limited to about twice the minimum beam size (at the waist) to ensure accuracy~\cite{houjunphdthesis}. As a result, the range of the solenoid strength $K$ during the scan is small compared to $k_0$. For example, the range only varies by 5.1\% during the solenoid scan introduced in Sec.~\ref{ex}. Therefore, the solenoid strength $K$ can be expanded about $k_0$
\begin{equation}\label{eq019}
K=k_0+\Delta k
\end{equation}
where $k_0$ is the solenoid strength that focuses the beam to the waist, and $\Delta k \ll k_0$ in the range. The relationship between $k_0$ and the drift length $L_d$ is given by
\begin{equation}\label{eq020}
{L_d} = \frac{{\cot (k_0L)}}{{k_0}}
\end{equation}
\par To obtain the Taylor series expansion of the screen beam spot size squared ($\sigma_{scr}^2$) we substitute Eqn.~\ref{eq019} and Eqn.~\ref{eq020} into Eqn.~\ref{eq100} give to second order in $\Delta k$ 
\begin{widetext}
\begin{equation}\label{eq32}
\begin{aligned}
{\sigma _{scr}}^2 =& \frac{\langle {x_{1}}^2\rangle}{{{f^2}{{k_0}^2}}}\left( {3{c^4} + {{{c^6}} \mathord{\left/
 {\vphantom {{{c^6}} {{s^2}}}} \right.
 \kern-\nulldelimiterspace} {{s^2}}} + 3{c^2}{s^2} + {s^4}} \right) + \frac{{2\langle {x_{1}}^2\rangle}}{{{f^2}{{k_0}^3}}}\left( \begin{array}{l}
f{{k_0}^2}L{c^4} - {c^4} + fk_0{{{c^5}} \mathord{\left/
 {\vphantom {{{c^5}} s}} \right.
 \kern-\nulldelimiterspace} s} + f{{k_0}^2}L{{{c^6}} \mathord{\left/
 {\vphantom {{{c^6}} {{s^2}}}} \right.
 \kern-\nulldelimiterspace} {{s^2}}} - \\
2{c^2}{s^2} - f{{k_0}^2}L{c^2}{s^2} - fk_0c{s^3} - {s^4} - f{{k_0}^2}L{s^4}
\end{array} \right)\Delta k\\
 &+
\frac{\langle {x_{1}}^2\rangle}{{{f^2}{{k_0}^4}}}\left( \begin{array}{l}
2{c^4} + {f^2}{{k_0}^2}{c^4} - 10f{{k_0}^2}L{c^4} - 3{{k_0}^2}{L^2}{c^4} + 3{f^2}{{k_0}^4}{L^2}{c^4} - 2k_0L{{{c^5}} \mathord{\left/
 {\vphantom {{{c^5}} s}} \right.
 \kern-\nulldelimiterspace} s} + 2{f^2}{{k_0}^3}L{{{c^5}} \mathord{\left/
 {\vphantom {{{c^5}} s}} \right.
 \kern-\nulldelimiterspace} s} - \\
8f{{k_0}^3}{L^2}{{{c^5}} \mathord{\left/
 {\vphantom {{{c^5}} s}} \right.
 \kern-\nulldelimiterspace} s} + 2f{{k_0}^2}L{{{c^6}} \mathord{\left/
 {\vphantom {{{c^6}} {{s^2}}}} \right.
 \kern-\nulldelimiterspace} {{s^2}}} - {{k_0}^2}{L^2}{{{c^6}} \mathord{\left/
 {\vphantom {{{c^6}} {{s^2}}}} \right.
 \kern-\nulldelimiterspace} {{s^2}}} + {f^2}{{k_0}^4}{L^2}{{{c^6}} \mathord{\left/
 {\vphantom {{{c^6}} {{s^2}}}} \right.
 \kern-\nulldelimiterspace} {{s^2}}} - 2fk_0{c^3}s - 4k_0L{c^3}s + \\
4{f^2}{{k_0}^3}L{c^3}s - 16f{{k_0}^3}{L^2}{c^3}s + 5{c^2}{s^2} + {f^2}{{k_0}^2}{c^2}{s^2} - 10f{{k_0}^2}L{c^2}{s^2} - 3{{k_0}^2}{L^2}{c^2}{s^2} + \\
3{f^2}{{k_0}^4}{L^2}{c^2}{s^2} + 2fk_0c{s^3} - 2k_0Lc{s^3} + 2{f^2}{{k_0}^3}Lc{s^3} - 8f{{k_0}^3}{L^2}c{s^3} + 3{s^4} + 2f{{k_0}^2}L{s^4} - \\
{{k_0}^2}{L^2}{s^4} + {f^2}{{k_0}^4}{L^2}{s^4}
\end{array} \right)(\Delta k)^{2}\\
 &+ {\rm O}((\Delta k)^{3})
\end{aligned}
\end{equation}
\end{widetext}
where $ s\equiv{\rm sin}(k_0L)$ and $c\equiv{\rm cos}(k_0L)$.

\par Similarly, to obtain the Taylor series expansion of the fitted beam spot size squared ($\sigma_{fit}^2$) we substitute Eqn.~\ref{eq019} and Eqn.~\ref{eq020} into Eqn.~\ref{eq33} 

\begin{widetext}
\begin{equation}\label{eq34}
\begin{aligned}
{\sigma _{fit}}^2 =& \frac{{\langle {{x'}_{fit}}^2\rangle }}{{{{k_0}^2}}}{\left( {{{{c^2}} \mathord{\left/
 {\vphantom {{{c^2}} s}} \right.
 \kern-\nulldelimiterspace} s} + s} \right)^2} - \frac{2}{{{{k_0}^3}}}\left( \begin{array}{l}
\langle {{x'}_{fit}}^2\rangle {c^2} + 2\langle {x_{fit}}{{x'}_{fit}}\rangle {{k_0}^2}L{c^2} + \langle {x_{fit}}{{x'}_{fit}}\rangle k_0{{{c^3}} \mathord{\left/
 {\vphantom {{{c^3}} s}} \right.
 \kern-\nulldelimiterspace} s} + \langle {x_{fit}}{{x'}_{fit}}\rangle {{k_0}^2}L{{{c^4}} \mathord{\left/
 {\vphantom {{{c^4}} {{s^2}}}} \right.
 \kern-\nulldelimiterspace} {{s^2}}} + \\
\langle {x_{fit}}{{x'}_{fit}}\rangle k_0cs + c{s^2} + \langle {x_{fit}}{{x'}_{fit}}\rangle {{k_0}^2}L{s^2}
\end{array} \right)\Delta k\\
 &+ 
\left( \begin{array}{l}
\frac{{\langle {x_{fit}}^2\rangle }}{{{{k_0}^2}}}{\left( {c + k_0L{{{c^2}} \mathord{\left/
 {\vphantom {{{c^2}} s}} \right.
 \kern-\nulldelimiterspace} s} + k_0Ls} \right)^2} - \frac{{2\langle {x_{fit}}{{x'}_{fit}}\rangle }}{{{{k_0}^3}}}\left( {k_0L{{{c^4}} \mathord{\left/
 {\vphantom {{{c^4}} {{s^2}}}} \right.
 \kern-\nulldelimiterspace} {{s^2}}} - cs - k_0L{s^2}} \right) + \\
\frac{{\langle {{x'}_{fit}}^2\rangle }}{{{{k_0}^4}}}\left( {{s^2} + \left( {{{{c^2}} \mathord{\left/
 {\vphantom {{{c^2}} s}} \right.
 \kern-\nulldelimiterspace} s} + s} \right)\left( {2s - 2k_0Lc - {{k_0}^2}{L^2}{{{c^2}} \mathord{\left/
 {\vphantom {{{c^2}} s}} \right.
 \kern-\nulldelimiterspace} s} - {{k_0}^2}{L^2}s} \right)} \right)
\end{array} \right)(\Delta k)^{2} + {\rm O}((\Delta k)^{3})
\end{aligned}
\end{equation}
\end{widetext}

\par By comparing of the coefficients of each $\Delta k$ power exponents in Eqn.~\ref{eq32} and Eqn.~\ref{eq34}, the fitted beam moments can be solved for as

\begin{widetext}
\begin{equation}\label{eq35}
\begin{array}{l}
\langle {x_{fit}}^2\rangle  = \frac{{\langle {x_{1}}^2\rangle\left( \begin{array}{l}
30{f^2}{{k_0}^3}L{c^8} - 160f{{k_0}^3}{L^2}{c^8} + 35{f^2}{{k_0}^5}{L^3}{c^8} + 4{f^2}{{k_0}^2}{{{c^9}} \mathord{\left/
 {\vphantom {{{c^9}} s}} \right.
 \kern-\nulldelimiterspace} s} - 32f{{k_0}^2}L{{{c^9}} \mathord{\left/
 {\vphantom {{{c^9}} s}} \right.
 \kern-\nulldelimiterspace} s} + 45{f^2}{{k_0}^4}{L^2}{{{c^9}} \mathord{\left/
 {\vphantom {{{c^9}} s}} \right.
 \kern-\nulldelimiterspace} s} - \\
120f{{k_0}^4}{L^3}{{{c^9}} \mathord{\left/
 {\vphantom {{{c^9}} s}} \right.
 \kern-\nulldelimiterspace} s} + 15{f^2}{{k_0}^3}L{{{c^{10}}} \mathord{\left/
 {\vphantom {{{c^{10}}} {{s^2}}}} \right.
 \kern-\nulldelimiterspace} {{s^2}}} - 80f{{k_0}^3}{L^2}{{{c^{10}}} \mathord{\left/
 {\vphantom {{{c^{10}}} {{s^2}}}} \right.
 \kern-\nulldelimiterspace} {{s^2}}} + 21{f^2}{{k_0}^5}{L^3}{{{c^{10}}} \mathord{\left/
 {\vphantom {{{c^{10}}} {{s^2}}}} \right.
 \kern-\nulldelimiterspace} {{s^2}}} + {f^2}{{k_0}^2}{{{c^{11}}} \mathord{\left/
 {\vphantom {{{c^{11}}} {{s^3}}}} \right.
 \kern-\nulldelimiterspace} {{s^3}}} - \\
8f{{k_0}^2}L{{{c^{11}}} \mathord{\left/
 {\vphantom {{{c^{11}}} {{s^3}}}} \right.
 \kern-\nulldelimiterspace} {{s^3}}} + 18{f^2}{{k_0}^4}{L^2}{{{c^{11}}} \mathord{\left/
 {\vphantom {{{c^{11}}} {{s^3}}}} \right.
 \kern-\nulldelimiterspace} {{s^3}}} - 48f{{k_0}^4}{L^3}{{{c^{11}}} \mathord{\left/
 {\vphantom {{{c^{11}}} {{s^3}}}} \right.
 \kern-\nulldelimiterspace} {{s^3}}} + 3{f^2}{{k_0}^3}L{{{c^{12}}} \mathord{\left/
 {\vphantom {{{c^{12}}} {{s^4}}}} \right.
 \kern-\nulldelimiterspace} {{s^4}}} - 16f{{k_0}^3}{L^2}{{{c^{12}}} \mathord{\left/
 {\vphantom {{{c^{12}}} {{s^4}}}} \right.
 \kern-\nulldelimiterspace} {{s^4}}} + \\
7{f^2}{{k_0}^5}{L^3}{{{c^{12}}} \mathord{\left/
 {\vphantom {{{c^{12}}} {{s^4}}}} \right.
 \kern-\nulldelimiterspace} {{s^4}}} + 3{f^2}{{k_0}^4}{L^2}{{{c^{13}}} \mathord{\left/
 {\vphantom {{{c^{13}}} {{s^5}}}} \right.
 \kern-\nulldelimiterspace} {{s^5}}} - 8f{{k_0}^4}{L^3}{{{c^{13}}} \mathord{\left/
 {\vphantom {{{c^{13}}} {{s^5}}}} \right.
 \kern-\nulldelimiterspace} {{s^5}}} + {f^2}{{k_0}^5}{L^3}{{{c^{14}}} \mathord{\left/
 {\vphantom {{{c^{14}}} {{s^6}}}} \right.
 \kern-\nulldelimiterspace} {{s^6}}} + 6{f^2}{{k_0}^2}{c^7}s - \\
48f{{k_0}^2}L{c^7}s + 60{f^2}{{k_0}^4}{L^2}{c^7}s - 160f{{k_0}^4}{L^3}{c^7}s + 30{f^2}{{k_0}^3}L{c^6}{s^2} - 160f{{k_0}^3}{L^2}{c^6}{s^2} + \\
35{f^2}{{k_0}^5}{L^3}{c^6}{s^2} + 4{f^2}{{k_0}^2}{c^5}{s^3} - 32f{{k_0}^2}L{c^5}{s^3} + 45{f^2}{{k_0}^4}{L^2}{c^5}{s^3} - 120f{{k_0}^4}{L^3}{c^5}{s^3} + 15{f^2}{{k_0}^3}L{c^4}{s^4}\\
 - 80f{{k_0}^3}{L^2}{c^4}{s^4} + 21{f^2}{{k_0}^5}{L^3}{c^4}{s^4} + {f^2}{{k_0}^2}{c^3}{s^5} - 8f{{k_0}^2}L{c^3}{s^5} + 18{f^2}{{k_0}^4}{L^2}{c^3}{s^5} - 48f{{k_0}^4}{L^3}{c^3}{s^5} + \\
3{f^2}{{k_0}^3}L{c^2}{s^6} - 16f{{k_0}^3}{L^2}{c^2}{s^6} + 7{f^2}{{k_0}^5}{L^3}{c^2}{s^6} + 3{f^2}{{k_0}^4}{L^2}c{s^7} - 8f{{k_0}^4}{L^3}c{s^7} + {f^2}{{k_0}^5}{L^3}{s^8}
\end{array} \right)}}{{{f^2}{{k_0}^2}{{\left( {{{{c^2}} \mathord{\left/
 {\vphantom {{{c^2}} s}} \right.
 \kern-\nulldelimiterspace} s} + s} \right)}^2}{{\left( {c + k_0L{{{c^2}} \mathord{\left/
 {\vphantom {{{c^2}} s}} \right.
 \kern-\nulldelimiterspace} s} + k_0Ls} \right)}^2}\left( {2k_0L{c^2} + {{{c^3}} \mathord{\left/
 {\vphantom {{{c^3}} s}} \right.
 \kern-\nulldelimiterspace} s} + k_0L{{{c^4}} \mathord{\left/
 {\vphantom {{{c^4}} {{s^2}}}} \right.
 \kern-\nulldelimiterspace} {{s^2}}} + cs + k_0L{s^2}} \right)}}\\
\langle {x_{fit}}{{x'}_{fit}}\rangle  = \frac{{{\sigma _{x0}}^2\left( {2f{{k_0}^2}L{c^6} + 2fk_0{{{c^7}} \mathord{\left/
 {\vphantom {{{c^7}} s}} \right.
 \kern-\nulldelimiterspace} s} + 3f{{k_0}^2}L{{{c^8}} \mathord{\left/
 {\vphantom {{{c^8}} {{s^2}}}} \right.
 \kern-\nulldelimiterspace} {{s^2}}} + fk_0{{{c^9}} \mathord{\left/
 {\vphantom {{{c^9}} {{s^3}}}} \right.
 \kern-\nulldelimiterspace} {{s^3}}} + f{{k_0}^2}L{{{c^{10}}} \mathord{\left/
 {\vphantom {{{c^{10}}} {{s^4}}}} \right.
 \kern-\nulldelimiterspace} {{s^4}}} - 2f{{k_0}^2}L{c^4}{s^2} - 2fk_0{c^3}{s^3} - 3f{{k_0}^2}L{c^2}{s^4} - fk_0c{s^5} - f{{k_0}^2}L{s^6}} \right)}}{{{f^2}k_0{{\left( {{{{c^2}} \mathord{\left/
 {\vphantom {{{c^2}} s}} \right.
 \kern-\nulldelimiterspace} s} + s} \right)}^2}\left( {2k_0L{c^2} + {{{c^3}} \mathord{\left/
 {\vphantom {{{c^3}} s}} \right.
 \kern-\nulldelimiterspace} s} + k_0L{{{c^4}} \mathord{\left/
 {\vphantom {{{c^4}} {{s^2}}}} \right.
 \kern-\nulldelimiterspace} {{s^2}}} + cs + k_0L{s^2}} \right)}}\\
\langle {{x'}_{fit}}^2\rangle {\rm{ = }}\frac{{{\sigma _{x0}}^2\left( {3{c^4} + {{{c^6}} \mathord{\left/
 {\vphantom {{{c^6}} {{s^2}}}} \right.
 \kern-\nulldelimiterspace} {{s^2}}} + 3{c^2}{s^2} + {s^4}} \right)}}{{{f^2}{{\left( {{{{c^2}} \mathord{\left/
 {\vphantom {{{c^2}} s}} \right.
 \kern-\nulldelimiterspace} s} + s} \right)}^2}}}
\end{array}
\end{equation}
\end{widetext}
\par Finally, using these fitted beam moments, we can calculate the measured (fitted) emittance at Position 1 as
\begin{equation}\label{eq33c}
\varepsilon_{fit}  = \sqrt {\langle {x_{fit}}^2\rangle \langle {{x'}_{fit}}^2\rangle  - {{\langle {x_{fit}}{{x'}_{fit}}\rangle }^2}} 
\end{equation}
which is, in general, not equal to actual emittance at Position 1, $\varepsilon _{1}=0$.
\par It is informative to calculate the actual emittance at the YAG screen (Position 3) for the beam waist condition.  Using Eqn.~\ref{eq31} when $K=k_0$ we find,
\begin{equation}\label{eq33b}
{\varepsilon _{3}} = \left| {{\Sigma _{scr}}(1:2,1:2)} \right| = \left| {\frac{{2\langle {x_{1}}^2\rangle c s}}{{f}}} \right|
\end{equation}
Note that this equation can also be derived with Eqn.~\ref{emittacne_coupled}.
\par
\begin{figure}[hbtp]
	\centering
	\includegraphics[width=0.4\textwidth]{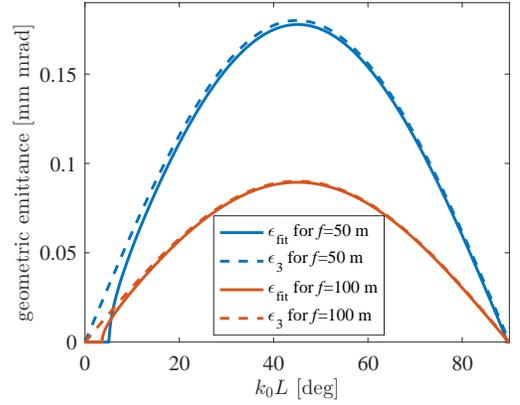}
	\caption{\label{FIG.emt_KL} Measured (fitted) emittance in the solenoid scan $\varepsilon_{fit}$ and the final emittance on the YAG screen $\epsilon_3$ as a function of Larmor angle and focal length.}
\end{figure}

 FIG.~\ref{FIG.emt_KL} compares the measured (fitted) emittance from the solenoid scan $\varepsilon_{fit}$ (Eqn.~\ref{eq33c}) to the final emittance at the end of the solenoid scan beamline $\varepsilon_{3}=\varepsilon_{coupled}$ (Eqn.~\ref{eq33b}) for the special case of zero initial emittance. Let $\sqrt{\langle {x_{1}}^2\rangle}$=3~mm and $L$=0.4~m, then $\epsilon_{fit}$ and $\epsilon_{3}$ are plotted as a function of the Larmor angle $k_0L$ and focal length $f$ in FIG.~\ref{FIG.emt_KL} and show good agreement. The difference between them is relatively large when the Larmor angle $k_0L$ is very small and $f$ is large, which are not common for realistic beamline parameters.  This plot shows that the measured (fitted) emittance $\varepsilon_{fit}$ is equal to the emittance after the solenoid beamline $\varepsilon _{3}$ for realistic beamline parameters.  In this specific case of zero initial emittance ($\varepsilon _{1}=0$), the final emittance is equal to emittance growth from the coupled aberration, $\varepsilon_{3}=\varepsilon_{coupled}$.   As we show next for the general solenoid scan case having $\varepsilon_{1}=\varepsilon_{therm}$, then the measured (fitted) emittance $\varepsilon_{fit}$ is still equal to the final emittance but it is now quadrature sum of $\varepsilon_{therm}$ and $\varepsilon_{coupled}$.
\par Next, the above analytic results for the solenoid scan are verified with numerical simulations using ASTRA but now will include a non-zero initial emittance. The beamline layout and the parameter settings used in this simulation are the same as in the experiment introduced in Sec.~\ref{ex}. The initial emittance is equal to the thermal emittance at the cathode, $\varepsilon_{0}=\varepsilon_{therm}$. A realistic laser intensity distribution was used (instead of a uniform distribution) in the ASTRA simulation leading to slight difference of the thermal emittance value between x and y planes. The CST 3D field map containing quadrupole field is used for the gun and a quadrupole component is added to the solenoid. The longitudinal field profile of the solenoid quadrupole is the same as the solenoid, and the solenoid quadrupole strength is proportional to the solenoid field strength.

\par In the ASTRA simulation of the solenoid scan measurement, the solenoid strength is scanned and a series of spot sizes at the screen are generated for the four different solenoid quadrupole settings shown in Table~\ref{t12}.  The emittance measured by the solenoid scan is calculated with the fitting method of Eqn.~\ref{eq33} and the ASTRA screen spot sizes. The simulation results are listed in Table~\ref{t12}, which include the initial emittance $\epsilon_{0}$, the actual emittance on the screen $\epsilon_{3}$, and the measured (fitted) emittance by solenoid scan $\epsilon_{fit}$. The variation of the actual emittance at the screen $\epsilon_{3}$ is due to the dependence of $\epsilon_{3}$ on the solenoid and solenoid quadrupole strength according to Eqn.~\ref{emittacne_coupled}. The results prove that the measured (fitted) emittance from the solenoid scan $\epsilon_{fit}$ is very close to the actual emittance at the screen $\epsilon_3$ (the quadrature sum of $\varepsilon_{therm}$ and $\varepsilon_{coupled}$). Therefore, the coupled transverse dynamics aberration can lead to an overestimation of the thermal emittance in the solenoid scan method.

\begin{table*}[hbtp]
\caption{\label{t12} Comparing the initial emittance on the cathode $\epsilon_{0}$, the actual emittance at the screen $\epsilon_{3}$, and the measured emittance with solenoid scan $\epsilon_{fit}$ under different strength and rotation angle of the solenoid quadrupole. The unit of the emittances is mm mrad.}
\renewcommand\tabcolsep{11pt}
\renewcommand\arraystretch{1.3}
\begin{tabular}{*{7}{c}}
\toprule[1.5pt]
                                     & $\epsilon_{x0}$   &  $\epsilon_{x3}$         & $\epsilon_{xfit}$    &  $\epsilon_{y0}$   &  $\epsilon_{y3}$        &  $\epsilon_{yfit}$    \\
\midrule[1pt]
0.005 T/m,$\eta$=0 deg      & 2.938 & 3.181-3.209 & 3.206  & 2.718 & 3.052-3.081 & 3.080  \\ 
0.005 T/m,$\eta$=45 deg    & 2.938 & 3.667-3.682 & 3.677 & 2.718 & 3.757-3.781 & 3.799 \\ 
0.01 T/m,$\eta$=0 deg       & 2.938 & 3.947-4.039 & 4.027  & 2.718 & 3.774-3.866 & 3.827  \\ 
0.01 T/m,$\eta$=45 deg     & 2.938 & 5.478-5.526 & 5.495 & 2.718 & 5.624-5.680 & 5.716 \\ 
\bottomrule[1.5pt]
\end{tabular}
\end{table*}

\section{thermal emittance measurement}\label{ex}
\par The thermal emittance of a cesium telluride photocathode was experimentally measured with the solenoid scan method at the Argonne Wakefield Accelerator (AWA) facility to experimentally demonstrate the overestimation of the thermal emittance due to the coupled transverse dynamics aberration. The layout of the beamline is shown in FIG.~\ref{FIG.beamline_show}. An L-band rf gun with a cesium telluride photocathode is illuminated by a 248~nm UV laser.  The transverse profile of the laser is homogenized with a micro-lens array~\cite{halavanau2017spatial}.  The cathode gradient is 32 MV/m and the laser launch phase with respect to rf is $43^\circ$ resulting in a beam energy of 3.2~MeV.  The bunch charge is kept below 1~pC to make the space charge effect negligible. A PI-MAX Intensified CCD (ICCD) camera~\cite{camera} with 100~ns shutter gating is employed to capture the beam images on the YAG screen. The resolution was measured to be 60 $\mu$m with a USAF target.
\subsection{thermal emittance in the linear regime}
\par
\begin{figure}[hbtp]
\centering
\includegraphics[width=0.4\textwidth]{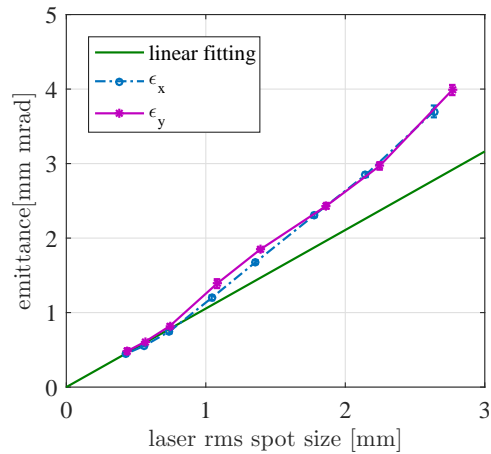}
\caption{\label{FIG.beamsize1} Measured emittance as a function of laser spot size. The slope of the line in the linear regime (laser spot size $<$0.75~mm) gives a thermal emittance of $1.05\pm0.04$ mm\,mrad/mm.}
\end{figure}

 The measured thermal emittance (i.e. the fitted emittance from the solenoid scan) for different laser spot sizes is shown in FIG.~\ref{FIG.beamsize1}.  The measured results can be classified into two regimes: a linear regime where the rms spot size is ($<$0.75~mm) and a nonlinear regime where the spot size is ($>$0.75~mm). In the linear regime, the measured emittance has a linear relationship with the spot size and the slope of this line ($1.05\pm0.04$ mm\,mrad/mm) can be used to extrapolate an accurate measurement of the thermal emittance from the fitted values.  This slope is in good agreement with the theoretical value of the thermal emittance of a cesium telluride photocathode illuminated by 248~nm laser as introduced in Sec.~\ref{section3}. In the nonlinear regime, the data deviates from the linear fit due to the coupled transverse dynamics aberrations. The deviation from the linear fit and the data is the overestimation of the thermal emittance and it becomes larger with increasing laser spot size, reaching 35\% at the largest laser rms spot size of 2.7~mm.
\par These results show that the thermal emittance can be measured with the solenoid scan method as long as one is able to extrapolate the measurements into the linear regime.  In our example of the AWA L-band (1.3 GHz) rf photoinjector we achieved an accurate measurement of the thermal emittance when the laser rms spot size is less than 0.75~mm. However, this becomes more difficult to do at higher frequency since the quadrupole fields in the gun and the solenoid become relatively stronger. Therefore, accurate measurements of the thermal emittance in rf photoinjectors are more difficult at S-band (2.856 GHz) and even more so for X-band (11.424 GHz).  If we assume that the size of the beamline elements scale as the inverse of the frequency, then the laser rms spot size should be less than 0.34~mm for S-band photoinjectors and less than 0.085~mm for X-band photoinjectors to eliminate the coupled transverse dynamics aberrations to get an accurate measurement of the thermal emittance using the solenoid scan.   From this we can conclude that lower frequency rf photoinjectors are preferred for accurate measurements of the thermal emittance. 

\subsection{Beam based method for the measurement of solenoid quadrupole in the nonlinear regime}
The strength and rotation angle of the effective solenoid quadrupole term can be inferred by using the solenoid scan in the nonlinear regime.  This method can be used in lieu of the measurement of the actual solenoid quadrupole term with a rotating wire when it is not practical to remove the solenoid from the beamline.  Note that the strength and rotation angle of the gun quadrupole are known from the 3D gun field map.

\begin{figure}[hbtp]
\centering
\includegraphics[width=0.48\textwidth]{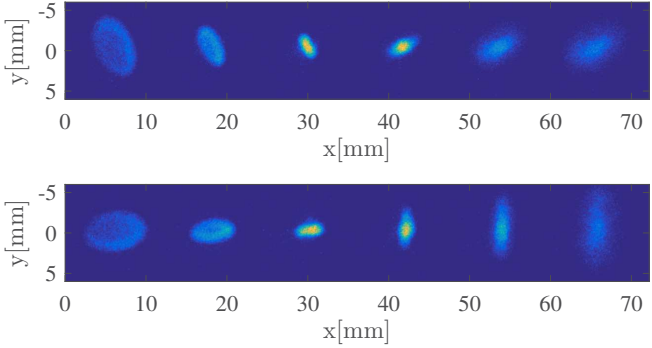}
\caption{\label{FIG.images} Electron beam images on the YAG screen for different solenoid currents. The upper and lower rows correspond to ccw and cw solenoid current directions respectively.}
\end{figure}

A large laser spot size (diameter=11~mm or rms=2.7~mm) is used in the following measurements to determine the solenoid quadrupole. Figure.~\ref{FIG.images} shows a series beam spots on the screen (Position 3) for various solenoid currents.  The top row corresponds to the solenoid current flowing in one direction while the current has been flipped in the bottom row (hereinafter these directions are referred to as counterclockwise (ccw) and clockwise (cw)).  The images of all the electron beams are elliptical for both the ccw and cw directions, however, the tilt angles are different for the two current directions.  Notice that the beam images in the top row are tilted and therefore have a strong x-y correlation while the bottom row has beams that are nearly normally oriented and therefore have a weak x-y correlation.  Since the coupled aberration is due to the x-y correlation we expect that when the solenoid current is in the cw direction (bottom row) the emittance growth should be less. This is indeed the case as we show below.

\par
\begin{figure}[hbtp]
\centering
\includegraphics[width=0.4\textwidth]{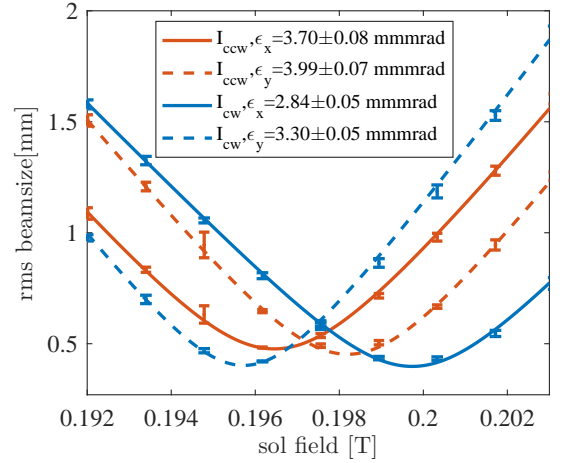}
\caption{\label{FIG.ex_result} Solenoid scan experiment data.  rms beamsize as a function of the solenoid field strength and fitted emittances taken during the solenoid scan experiment.}
\end{figure}

 Figure~\ref{FIG.ex_result} shows the $x$ and $y$ rms beam spot size as a function of the solenoid strength during the scan for both directions of the solenoid current.  While the theoretical thermal emittance is 2.835~mm\,mrad, both solenoid scan fits yield higher emittance values as expected from the coupled aberration.  Note that the emittance overestimation is larger for the ccw direction than the cw direction; as expected.

\par
\begin{figure}[hbtp]
\centering
\includegraphics[width=0.4\textwidth]{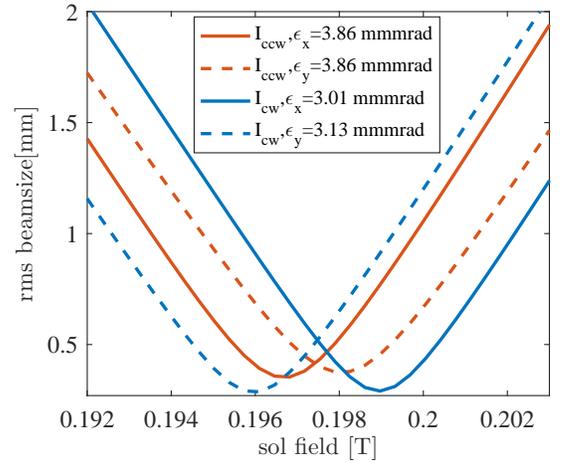}
\caption{\label{FIG.ex_result_si} Solenoid scan simulation fit results. rms beam size as a function of the solenoid field in ASTRA. The solenoid quadrupole was scanned to fit the simulation to the data.}
\end{figure}

 To estimate the strength and rotation angle of the effective solenoid quadrupole an ASTRA simulation of the experimental results is performed.   The simulation uses the 3D rf gun field map (therefore the gun quadrupole is assumed known), an ideal solenoid, and an ASTRA quad element to model solenoid quadrupole.  The ASTRA quadrupole element has: (i) length equal to the solenoid and (ii) strength proportional to the solenoid current.  Two variables of the quadrupole element, its strength and rotation angle, are numerically scanned to fit the simulation to the experimental results of Figure~\ref{FIG.ex_result}.  The best fit of the simulation is shown in FIG.~\ref{FIG.ex_result_si}.

\par
\begin{figure}[hbtp]
\centering
\includegraphics[width=0.43\textwidth]{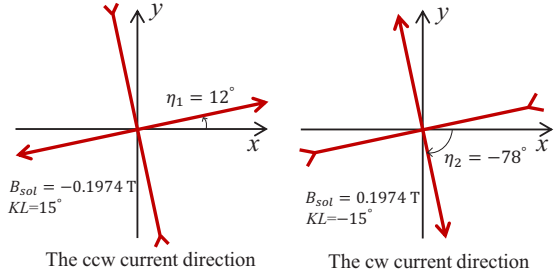}
\caption{\label{FIG.solenoid_reverse_demo} Left: The solenoid field is in the $-z$ direction for the ccw solenoid current direction, and the rotation angle of the solenoid quadrupole is $12^\circ$. Right: The solenoid field is in the $+z$ direction for the cw solenoid current direction, and the rotation angle of the solenoid quadrupole becomes $-78^\circ$. The rotation angle is defined as the angle between the quadrupole focusing direction and the x-axis.}
\end{figure}

 The fit of the ASTRA simulation to the experimental results yields the strength of the quadrupole solenoid as 77~Gauss/m for a solenoid field of 0.1974~T. The rotation angles of the quadrupole solenoid are different for ccw and cw current directions.  The values of the rotation angles are illustrated in FIG.~\ref{FIG.solenoid_reverse_demo}. For the ccw current direction, the solenoid field is -0.1974~T, the corresponding Larmor angle is $15^\circ$, and the rotation angle of the quadrupole is $12^\circ$. For the cw current direction, the solenoid field is 0.1974~T, the corresponding Larmor angle is $-15^\circ$, and the rotation angle of the quadrupole is $-78^\circ$. The emittance growth contributed by the gun quadrupole is constant regardless of the change of the solenoid current direction, so the difference of the emittances between the two solenoid current directions is determined by the quadrupole in the solenoid. According to Eqn.~\ref{emittacne_coupled}, the emittance growth due to the solenoid quadrupole is proportional to $\left| {\sin 2(KL + \eta )} \right|$. For the parameters shown in FIG.~\ref{FIG.solenoid_reverse_demo}, $\left| {\sin 2(KL + \eta )} \right|=0.809$ and $0.1045$ for the ccw and the cw current direction respectively, which explains the larger emittance overestimation for the ccw current direction.  In summary, this beam based method shows that the AWA solenoid quadrupole has strength 77~Gauss/m and rotation angle $12^\circ$ for the ccw direction and $-78^\circ$ for the cw direction.

\section{quadrupole corrector}\label{corrector}

\par
\begin{figure}[hbtp]
\centering
\includegraphics[width=0.43\textwidth]{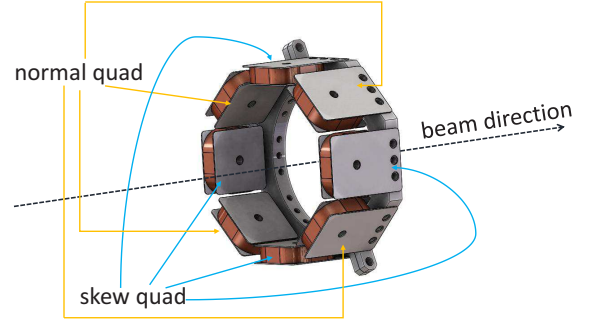}
\caption{\label{FIG.corrector} The mechanical design of the quadrupole corrector at AWA.}
\end{figure}

 To eliminate the emittance growth due to the coupled transverse dynamics aberration, several types of quadrupole correctors have been proposed.  A quadrupole corrector is useful for the specific case of improving the solenoid scan fidelity but also for the more general case of reducing emittance growth \cite{dowell2018exact,bartnik2015operational,schietinger2016commissioning,krasilnikov2018electron}.  A dedicated quadrupole corrector has also been designed for use at the AWA to cancel the emittance overestimation in solenoid scan, as shown in FIG.~\ref{FIG.corrector}. The corrector consists of a pair of normal and skew quadrupoles in order to obtain a quadrupole with variable strength and rotation angle.

\begin{figure}[hbtp]
\centering
\includegraphics[width=0.43\textwidth]{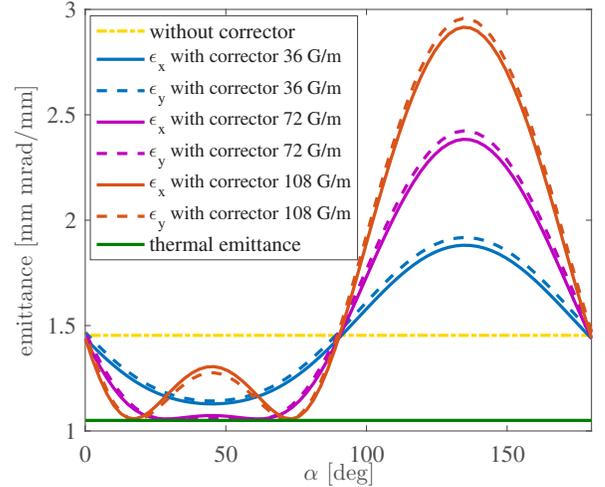}
\caption{\label{FIG.emt_theta_paper1} Simulated emittance correction by scanning the strength and rotation angle $\alpha$ of the quadrupole corrector. The emittance is normalized by the laser spot size.}
\end{figure}

The quadrupole corrector will be installed at the solenoid exit. An ASTRA simulation is employed to study the correction effect. The solenoid quadrupole used in this simulation is in the ccw direction as described in Sec.~\ref{ex}. The laser spot is uniform in the transverse direction with a rms spot size of 2.7~mm. The other parameters are kept the same as in the experiment. Since the scan range of the solenoid strength is small in the solenoid scan (5.1\% of the solenoid strength), a constant strength of the quadrupole corrector is sufficient to cancel the emittance growth. Figure~\ref{FIG.emt_theta_paper1} shows the measured emittance as a function of the quadrupole corrector strength and rotation angle. The figure clearly shows that the amount of the emittance overestimation due to the coupled transverse dynamics aberration depends on the strength and rotation angle of the quadrupole corrector.  Moreover, the emittance measured by the solenoid scan can be made equal to the thermal emittance if the quadrupole corrector setting is chosen appropriately.

\section{conclusion}

\par The overestimation of the thermal emittance due to the coupled transverse dynamics aberration in solenoid scan has been systematically studied in this paper. Two sources of aberrations that lead to emittance growth were analyzed: the quadrupole field in the rf gun followed by a solenoid, and the quadrupole field of the solenoid. Analytical expressions and beam dynamics simulations demonstrated that the emittance measured by solenoid scan is an overestimation which is very close to the quadrature sum of the thermal emittance and coupled emittance. 
\par The overestimation effect in the solenoid scan was demonstrated with a thermal emittance measurement experiment at the AWA facility.  The experiment measured the thermal emittance of a cesium telluride photocathode in the AWA drive gun, as L-band 1.6-cell rf gun. Elliptical beam images were observed on the YAG screen indicating the existence of quadrupole components in the beamline. A nonlinear curve of the measured emittance as a function of the laser spot size is observed, which shows a thermal emittance overestimation of 35\% with a 2.7 mm rms laser spot size. A beam based method was used to  measure the solenoid quadrupole by flipping the solenoid current direction and matching the simulation with the experimental results. Its strength is found to be 77 Gauss/m with a solenoid field of 0.1974 T and its rotation angle is discovered to be $12^{\circ}$ and $-78^{\circ}$ with the two opposite solenoid current directions, respectively.
\par A compact and flexible quadrupole corrector is proposed to be installed at the exit of the solenoid, which will fully eliminate the overestimation effect due to the coupled transverse dynamics aberrations so as to improve the thermal emittance measurement accuracy by the solenoid scan method.

\begin{acknowledgments}
  \par  This work is supported by the U.S. Department of Energy, Offices of HEP and BES, under Contract No. DE-AC02-06CH11357. It is also funded by the National Natural Science Foundation of China (NSFC) No. 11435015 and No. 11375097.
\end{acknowledgments}

% Create the reference section using BibTeX:
\bibliography{apstemplate.bib}

\end{document}